\documentclass{iaus}
\usepackage{graphicx}

\title[The LCID project: data reduction strategy] 
{The ACS LCID project: data reduction strategy}

\author[M. Monelli et al.]   
{M. Monelli$^1$ for the LCID team\thanks{Local Cosmology from Isolated Dwarfs, http://www.iac.es/project/LCID}}

\affiliation{$^1$Insituto de Astrof\'isica de Canarias, C/ Via Lactea s/n, 38205 La Laguna,
Tenerife, Spain \break email: monelli@iac.es}

\pubyear{2007}
\volume{241}  
\pagerange{119--126}
\date{?? and in revised form ??}
\jname{Stellar Populations as Building Blocks of Galaxies}
\editors{A. Vazdekis \& R. Peletier, eds.}
\begin{document}

\maketitle

\begin{abstract}
During Cycle 14 a total of 113 HST orbits were secured to observe 
five isolated dwarf galaxies, namely Tucana, LGS3, LeoA, IC1613, and 
Cetus. The aim of the project is a full characterization of the stellar 
content of these galaxies, in term of their SFH, radial distributions, halo populations 
and variable stars. Deep (V$\approx$29) F475W, F814W data allowed us to fully sample all the 
evolutionary phases from the tip of the Red Giant Branch (RGB) to well below 
the old Main Sequence Turnoff (MSTO). Here we describe the observational design, 
and the reduction and calibration strategy adopted. A comparison of the results obtained using 
two different packages, ALLFRAME and Dolphot, is presented.
\end{abstract}

The main goal of this project (Gallart et al., this proceedings) is to obtain 
deep color-magnitude diagrams (CMDs) reaching the oldest MSTO (M$_I \simeq +3$),
in order to obtain detailed SFHs and to perform 
comparative studies within the sample and with available data for Milky Way satellites. 
The SFH will be recovered from the global analysis of the star distribution 
on the CMD, using IAC-star and IAC-pop (Aparicio et al., this proceedings).
The observational design was determined as follows.
The most challenging part of our project was to trace {\it age differences} at old ages,
and we found that the largest age sensitivity of old isochrones occurs 
as color differences at $M_I\simeq +2.75$. We also determined that the F475W and F814W 
filters is the most efficient combination to trace it.
We calculated that, in order to trace age differences
of about 2 Gyr at the above magnitude level, a precision of $\simeq
0.1$ magnitudes was necessary. This determined the number of orbits
devoted to each object. Moreover, these exposure
times allow us to reach M$_I \simeq +2.5$ with S/N $\simeq$ 10 in the
parallel WFPC2 exposures, using the F450W and F814W filters. The last are
expected to produce CMDs showing stars 
down to $\simeq$ 5 Gyr on the MS, but with the old MSTOs below 
the limiting magnitude. The WFPC2 parallels were strategically 
placed in order to sample the outer parts of each galaxy, 
and thus to provide information on stellar population gradients. 
In the case of the smallest objects, they may provide limits to the 
extent of the galaxy.

The photometry of these data is being performed in parallel using two different
software packages, namely DAOPHOT/ALLFRAME (\cite[Stetson 1994]{steston-allf})
and Dolphot, an adaptation of HSTphot (\cite[Dolphin 2000]{dolphin00}) to
the perform ACS photometry.
This will allow us to assess the possible systematic effects on the CMD,
which may affect the SFH derivation.
We adopted similar approaches with both packages. After many tests,
we decided to work with the original $\_FLT$ images rather the drizzled ones.
The PSF was modeled individually for each image, and the input list
of stars was obtained from the cosmic-ray-cleaned stacked median image.
The photometric calibration was performed in the VEGAMAG system, following
the prescriptions given by \cite{sirianni05}.
Fig. 1 (left) shows the comparison between the Tucana CMD
obtained with the two methods. The right panels show, as
a function of the $F475W$ magnitude, the $\Delta(F475W)$, $\Delta(F814)$,
and $\Delta(F475W-F814W)$. The comparison discloses a possible small zero point and
a systematic trend towards faint magnitudes, being ALLFRAME magnitudes fainter then 
Dolphot ones. However, this trend appears to be similar in both bands, and
appears not to significantly affect the color. The differences between the results
of the two packages are small, and this should not affect the derived SFH in any way.
To qualitatively show this, theoretical isochrones from the BASTI database 
(\cite[Pietrinferni \etal\ 2004]{pietrinferni04}, 
http://www.te.astro.it/BASTI/index.php) are superimposed on the CMDs.
Extensive experiments are being performing to set this point on a quantative basis.

\begin{figure}
 \includegraphics[height=8.1truecm]{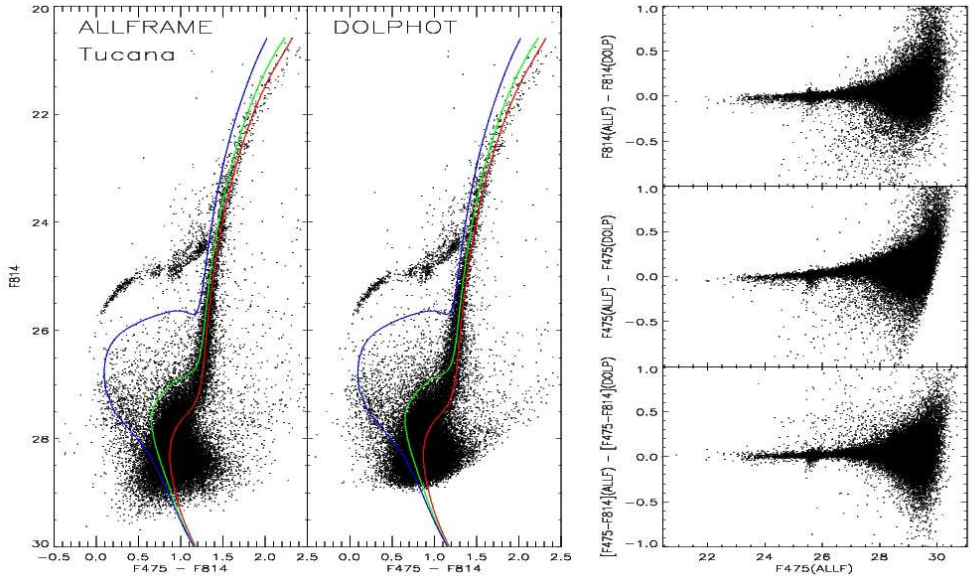}
  \caption{{\it Left -} Allframe and Dolphot CMDs. The same 
  theoretical isocrones (Z=0.006, t=11, 7, 3.5 Gyr, DM$_0$=24.7, E(B-V)=0.03) from the 
  BASTI database have been superimposed on both plots. {\it Right - } Magnitude and
  color differences as a funcion of $F475W$ magnitude.
  }\label{fig:fig}
\end{figure}

\begin{acknowledgments}
This work was partially supported by spanish MEC (AYA2004-06343)
the IAC (3I0394), by the European Structural Funds and NASA through grants 
GO-10505 and GO-10590
\end{acknowledgments}

\end{document}